\documentclass[11pt,a4paper]{article}
\usepackage{jcappub}
\usepackage{mathrsfs}
\usepackage{epsfig}
\usepackage{cite}
\usepackage{amssymb}
\usepackage{rotating}

\usepackage{amsmath}
\usepackage{amssymb}
\usepackage{graphicx}
\usepackage{color}

\title{Bias-Limited Extraction of Cosmological Parameters}

\author[1]{Meir Shimon,}\emailAdd{meirs@wise.tau.ac.il}
\author[1]{Nissan Itzhaki,}\emailAdd{nitzhaki@post.tau.ac.il}
\author[1,2]{Yoel Rephaeli}\emailAdd{yoelr@wise.tau.ac.il}
\affiliation[1]{School of Physics and Astronomy, Tel Aviv University, Tel Aviv
69978, Israel}
\affiliation[2]{Center for Astrophysics and Space Sciences, University of California,
San Diego, La Jolla, CA, 92093}

\author{}

\abstract{It is known that modeling uncertainties and astrophysical foregrounds 
can potentially introduce appreciable bias in the deduced values of cosmological 
parameters. While it is commonly assumed that these uncertainties will be 
accounted for to a sufficient level of precision, the level of bias has not been 
properly quantified in most cases of interest. We show that the requirement that 
the bias in derived values of cosmological parameters does not surpass nominal 
statistical error, translates into a maximal level of {\it overall} 
error $O(N^{-\frac{1}{2}})$ on $|\Delta P(k)|/P(k)$ and $|\Delta C_{l}|/C_{l}$, 
where $P(k)$, $C_{l}$, and $N$ are the matter power spectrum, angular power 
spectrum, and number of (independent Fourier) modes at a given scale $l$ or $k$ 
probed by the cosmological survey, respectively. This required level has important 
consequences on the precision with which cosmological parameters are hoped to be 
determined by future surveys: In virtually all ongoing and near future surveys $N$ 
typically falls in the range $10^{6}-10^{9}$, implying that the required overall 
theoretical modeling and numerical precision is already very high. Future 
redshifted-21-cm observations, projected to sample $\sim 10^{14}$ modes, will 
require knowledge of the matter power spectrum to a fantastic $10^{-7}$ precision 
level. We conclude that realizing the expected potential of future cosmological 
surveys, which aim at detecting $10^{6}-10^{14}$ modes, sets the formidable 
challenge of reducing the overall level of uncertainty to $10^{-3}-10^{-7}$.} 


\begin{document}

\maketitle

\flushbottom

\section{Introduction}

The last two decades marked the beginning of the `precision cosmology' era when many 
cosmological surveys are conducted in order to determine about a dozen cosmological 
parameters, and test non-standard cosmological models. 
These cosmological surveys are motivated by detailed theoretical predictions, 
elaborately designed satellite and stratospheirc telescope systems, and 
computationally challenging data analysis and parameter inference procedures. 

It is often stated that the joint effort of a variety of cosmological probes will 
enable breaking some of the vexing degeneracies in cosmological parameter space, 
but perhaps more importantly, it is hoped that cross-correlating the results 
obtained with a battery of cosmological probes could be used to mitigate 
inconsistencies due to the different systematics affecting each probe.

In this work we generalize basic arguments, first discussed by Seljak et al. [1] 
in the context of assessing the degree of agreement between different Boltzmann 
codes employed to calculate power spectra of the cosmic microwave background (CMB). 
We argue that as cosmological surveys become more advanced by virtue of higher 
spectro-spatial resolution and sensitivity, larger volume coverage, and lower 
instrumental noise, the need for a better understanding of the theory, higher 
numerical precision, a model-independent description of nonlinear effects, 
astrophysical foregrounds, and instrumental systematics, are all essential for 
achieving the full benefit of these advanced cosmological probes. 
The {\it quantitative} implications and ensuing ramifications of this (quite 
intuitive) statement are of primary interest for this work.

The challenge stems from the fact that the statistical error in inferred 
cosmological parameters decreases as $\sim N^{-1/2}$, where $N$ is the number of 
independent Fourier modes that can be probed by a given experiment. In contrast, 
the bias (induced by e.g. inaccurate theory or model, systematics, and foreground 
removal) does not decrease with the number of modes. The problem then arises with 
those advanced probes that target a large number of modes where the tension 
between the two trends becomes sufficiently large to result in an unavoidable 
bias in the inferred cosmological parameters. 

Current cosmological probes access $O(10^{6})-O(10^{9})$ modes, whereas next 
generation experiments aim at probing over $10^{14}$ modes, most notably 
redshifted-21-cm observations, which allows 1-2 orders of magnitude tighter 
constraints to be placed on a few cosmological parameters (e.g. [2-5]). It is 
clear that either an unimaginable precision in theoretical modeling has to be 
achieved, or the number of modes used for parameter estimation must be cut at 
lower angular (2D) or spatial (3D) resolution than previously projected, thereby 
significantly degrading the scientific yield of these probes with respect to 
standard estimates.

The outline of the paper is as follows. In section 2 we provide a general 
discussion of biased parameter inference in the framework of Fisher matrix 
formalism, followed by applications to angular and matter power spectra. 
In section 3 we discuss a few specific examples of systematics and modeling 
uncertainties and highlight their relevance to biased parameter inference. 
Our conclusions are summarized in section 4.

\section{Statistical Error and Bias}

Analyses of cosmological surveys, such as galaxy correlations, galaxy shear, CMB, 
and supernovae (SNIe) are commonly based on maximum likelihood methods to obtain 
the best fit cosmological model to the data. To assess bias in cosmological 
parameter extraction, we assume the likelihood function is gaussian in the power 
spectra. Specifically, this function is gaussian in the angular CMB power spectra, 
in the matter power spectrum of large scale structure (LSS) proxies (e.g. galaxy 
clustering, galaxy lensing, BAO, Ly$\alpha$, 21cm), and in the luminosity distance 
of SNe. While this function is poissonian in number counts of galaxy clusters, the 
numbers are sufficiently large to warrant a gaussian approximation.

In the framework of standard cosmology, it is clear that its inherent degree of 
symmetry (either the global 2D sky isotropy or 3D spatial homogeneity of the 
universe) implies that the universe is densely sampled on small angular (2D) and 
physical scales (3D), and thereby limited only by the fundamental minimum scale 
that could be accessed by a specific statistical probe with a given experiment 
(i.e. by the multipole number $l$ and wavenumber $k$ in the 2D and 3D cases, 
respectively, rather than the 2D and 3D wave-vectors ${\bf l}$ and ${\bf k}$). 
In the ideal case of no instrumental noise this implies that the only fundamental 
limit on precision is the so-called `cosmic-variance-limit' (CVL), i.e. the fact 
that we observe only a single realization of the universe out of an infinitely 
large number of possible universes characterized by the same model (with all 
models having the exact same variance, i.e. power spectrum, but different one-point 
functions). 

We begin quantitative assessment of statistical error and bias by a brief derivation 
of the well-known expressions for these measures within the Fisher matrix formalism. 
We then apply this formalism specifically to the angular and matter power spectra.

Without limiting the generality of the discussion we assume the likelihood function is 
gaussian in the data, at least in the large numbers limit. For a quantity $d_{n}$ for 
which there are $n_{max}$ data points $\hat{d}_{n}$, where $n=1, ... ,n_{max}$ stands 
for the multipole number $l$ in the case of CMB and galaxy lensing, $k$ and $z$ in 
the case of large scale proxies, and redshift bin in the case of SNe and number counts, 
the likelihood function can then be written as 
\begin{eqnarray}
\mathcal{L}=\exp\left[-\sum_{n=1}^{n_{max}}\frac{(d_{n}
-\hat{d}_{n})^{2}}{2(\delta d_{n})^{2}}\right]
\end{eqnarray}
where $\delta d_{n}$ is the statistical uncertainty on $\hat{d}_{n}$ due to, e.g., 
observational, cosmic variance, and shot noise (in the case of galaxy lensing). 
Let us assume that there are $M$ model parameters, 
$\boldsymbol\lambda=(\lambda_{1},...,\lambda_{M})$. Taylor expansion of $d_{n}$ 
around the best fit model, $\boldsymbol\lambda_{0}$, keeping terms to first order, 
we obtain
\begin{eqnarray}
d_{n}(\boldsymbol\lambda)\approx d_{n}(\boldsymbol\lambda_{0})
+\frac{d(d_{n})}{d\boldsymbol\lambda}
\cdot(\boldsymbol\lambda-\boldsymbol\lambda_{0}), 
\end{eqnarray}
where we use boldface letters for either vectors or 2D matrices. The likelihood 
function, Eq.(2.1), is then,  
\begin{eqnarray}
\mathcal{L}\approx\exp\left[-\frac{1}{2}\sum_{n=1}^{n_{max}}
(\boldsymbol\lambda-\boldsymbol\lambda_{0})\cdot{\bf F}_{n}
\cdot(\boldsymbol\lambda-\boldsymbol\lambda_{0})\right]
\end{eqnarray}
where for a given $n$ we have defined
\begin{eqnarray}
({\bf F}_{n})_{ij}\equiv\frac{1}{(\delta d_{n})^{2}}
\frac{d(d_{n})}{d\lambda_{i}}\frac{d(d_{n})}{d\lambda_{j}} 
\end{eqnarray}
and $(F)_{ij}=\sum_{n}({\bf F}_{n})_{ij}$ is the $ij$ element of the familiar 
Fisher matrix. The statistical error on the parameter $\lambda_{i}$ is
\begin{eqnarray}
\sigma_{\lambda_{i}}=\sqrt{(({\bf F})^{-1})_{ii}}.
\end{eqnarray}

Assume that we have a modeling error, or numerical imprecision, and that our 
theoretical model shifts by $\Delta d_{n}$, 
i.e. $d_{n}\rightarrow d_{n}+\Delta d_{n}$, 
that is either unknown or otherwise cannot be quantified and accounted for. 
In the following we consider a single data point for notational simplicity, 
for which the likelihood function is 
\begin{eqnarray}
\mathcal{L}\rightarrow\mathcal{L}'=\exp\left[-\frac{[\Delta d
+\frac{d(d)}{d\boldsymbol\lambda}\cdot(\boldsymbol\lambda
-\boldsymbol\lambda_{0})]^{2}}{2(\delta d)^{2}}\right].
\end{eqnarray}
Solving for the peak of the likelihood function, we obtain the shift, 
i.e. bias, in the best-fit parameters
\begin{eqnarray}
\delta_{\boldsymbol\lambda}\equiv \boldsymbol\lambda-\boldsymbol\lambda_{0}=
\frac{\Delta d}{(\delta d)^{2}}{\bf F}^{-1}\cdot\frac{d(d)}{d\boldsymbol\lambda}.
\end{eqnarray}

Now, assuming the bias and statistical error are uncorrelated we can add the 
nominal statistical error (Eq. 2.5) and the bias (Eq. 2.7) in quadrature
\begin{eqnarray}
\sigma_{\lambda_{i}}^{2}\rightarrow \sigma_{\lambda_{i}}^{2}
+(\delta_{\lambda_{i}})^{2}.
\end{eqnarray}

Comparing Eqs.(2.5) \& (2.7) we see that in order for the model bias $\Delta d$ 
to generate a significant parameter bias it has to be at the level of (or larger 
than) the variation of the theoretical $d$ over the `allowed' range of parameter 
values $\Delta d\gtrsim\frac{d(d)}{d\lambda}\sigma_{\lambda}$. In other words, 
while statistical error drops with increasing number of modes, any systematic 
bias in the theoretical model does not; therefore, for a given theoretical or 
modeling accuracy, there is a maximal mode beyond which the bias exceeds the 
statistical error. This implies that even if $\Delta d$ is smaller than $\hat{d}$ 
by a very large factor it can still result in a much more significant bias in 
parameter inference if the sensitivity of $d$ to small variations in the 
cosmological model is large. In that case the bias in the data, $\Delta d$, can 
mimic parameter shift, resulting in a biased parameter inference.

\subsection{CMB and Matter Power Spectra}

To illustrate our basic argument in the context of the CMB it is sufficient to 
consider the temperature-only dataset. In the absence of instrumental or any 
other noise source we have
\begin{eqnarray}
\delta C_{l}=\sqrt{\frac{1}{(2l+1)f_{sky}}}C_{l},
\end{eqnarray}
which is the cosmic variance. Note that as the sky fraction, $f_{sky}$, is 
smaller, sample variance, $\delta C_{l}$, increases. In the presence of 
instrumental noise this expression is modified to
\begin{eqnarray}
\delta C_{l}=\sqrt{\frac{1}{(2l+1)f_{sky}}}(C_{l}+C_{l}^{noise})
\end{eqnarray}
where $C_{l}^{noise}$ is the noise power spectrum. Typically, the detector 
(instrumental) noise sets a natural angular cutoff due to the final beam size,
\begin{eqnarray}
C_{l}^{noise}=(\Delta_{T}\theta_{b})^{2}e^{l(l+1)\sigma_{b}^{2}}
\end{eqnarray}
where $\theta_{b}=\sqrt{(8\ln(2)}\sigma_{b}$ is the beam full width at half 
maximum (FWHM), and $\Delta_{T}$ characterizes the detector's white noise 
level in temperature units. Power at multipoles $l>(\sigma_{b})^{-1}$ is 
exponentially downweighted. In other words, the experiment is 
cosmic-variance-limited only for $C_{l}\lesssim C_{l}^{noise}$.

In this CMB case Eqs.(2.3)-(2.5) yield 
\begin{eqnarray}
\mathcal{L}=\exp\left[-\frac{(\lambda-\lambda_{0})^{2}}{2
\sigma_{\lambda}^{2}}\right]
\end{eqnarray}
where
\begin{eqnarray}
\sigma_{\lambda}=\left[\sum_{l}f_{sky}\left(\frac{2l+1}{2}\right)
\frac{(\partial C_{l}/\partial\lambda)^{2}}{(C_{l}^{tot})^{2}}
\right]^{-1/2}
\end{eqnarray}
is the standard devitation of the parameter $\lambda$, i.e. its $1\sigma$ 
statistical uncertainty, and for simplicity we define
\begin{eqnarray}
C_{l}^{tot}\equiv C_{l}+C_{l}^{noise}.
\end{eqnarray}
Now, the bias in the parameter $\lambda$ in the presence of 
unaccounted-for contribution to the observed power spectrum, i.e. 
$C_{l}^{tot}\rightarrow C_{l}^{tot}+\Delta C_{l}$, is easily derived 
from Eq. (2.7). We attempt at fitting the observed power spectra with 
the `wrong' theoretical model. For example, we fit the primordial 
CMB `contaminated' by cluster- or filament-induced temperature anisotropy, 
the thermal Sunyaev-Zeldovich (SZ) effect, with theoretical model that 
accounts for the CMB only, without inclusion of filament-induced power. 
Doing so will clearly result in shifting (biasing) the best-fit $\lambda$, 
i.e. $\lambda_{0}\rightarrow\lambda_{0}+\delta_{\lambda}$. Of interest is 
the dimensionless bias, i.e. the bias in units of nominal statistical 
uncertainty, which gauges the bias importance; 
$\frac{\delta_{\lambda}}{\sigma_{\lambda}}>1$ indiactes a relatively large 
bias, whereas the bias is small if  
$\frac{\delta_{\lambda}}{\sigma_{\lambda}}<1$. 
Following a similar procedure to Eqs.(2.6)-(2.8), when there is an 
unaccounted contribution to the power spectrum, the likelihood function is 
\begin{eqnarray}
\mathcal{L}=\exp\left[-\sum_{l}\frac{[\Delta C_{l}
+(\frac{\partial C_{l}}{\partial\lambda})(\lambda
-\lambda_{0})^{2})]^{2}}{2(\delta C_{l})^{2}}\right]
\end{eqnarray} 
and the bias in the parameter $\lambda$ is 
\begin{eqnarray}
\lambda\sigma_{\lambda}^{-2}=\lambda_{0}\sigma_{\lambda}^{-2}
-f_{sky}\sum_{l}\left(\frac{2l+1}{2}\right)
\frac{\frac{\partial C_{l}}{\partial\lambda}\Delta C_{l}}{(C_{l}^{tot})^{2}}.
\end{eqnarray}
Therefore, the dimensionless bias reads
\begin{eqnarray}
\frac{\delta_{\lambda}}{\sigma_{\lambda}}=-f_{sky}\sigma_{\lambda}
\sum_{l}\left(\frac{2l+1}{2}\right)\frac{\frac{\partial C_{l}}{\partial
\lambda}\Delta C_{l}}{(C_{l}^{tot})^{2}}.
\end{eqnarray}
Biased cosmological parameter inference from CMB probes have already been 
discussed in the context of patchy reionization models (e.g. [6-7]) 
and residual `contamination' of the CMB sky by 
undetected galaxy clusters [8-9].

It is constructive at this point to consider a toy model that will provide 
an order of magnitude estimate of how large can the bias be. 
Assume for this toy model only that both $C_{l}$, 
$\frac{\partial C_{l}}{\partial\lambda}$ 
and $\Delta C_{l}$ are independent of $l$. In this case, Eqs.(2.13) 
\& (2.17) give 
\begin{eqnarray}
\sigma_{\lambda}&=&\left(\frac{f_{sky}}{2}l_{max}^{2}\right)^{-1/2}
\left(\frac{\partial C/\partial\lambda}{C}\right)^{-1}\nonumber\\
\frac{\delta_{\lambda}}{\sigma_{\lambda}}&=&\sqrt{\frac{f_{sky}}{2}}
l_{m}\frac{\Delta C}{C}
\end{eqnarray}
where here $l_{m}$ is the effective maximum $l$. 

For PLANCK and a CVL experiment, $l_{m}\approx 2500-3000$ and $f_{sky}\sim 1$. 
This implies that for a fixed $\delta_{\lambda}/\sigma_{\lambda}\approx 1$ the 
requirement is that $|\Delta C|/C\lesssim 1/l$, and that even if $|\Delta C|/C$ 
is at the 0.1\% level, we expect the dimensionless bias to be of order unity, 
i.e. it starts competing with the statistical uncertainty. Clearly, all power 
spectra in our case are $l$-dependent and this toy model does not apply. 
Nevertheless, it illustrates that while the large number of $l$-modes decreases 
the statistical uncertainty (as $l_{m}^{-1}$), the bias is hardly affected. 
As a result, the dimensionless bias scales as $l_{m}$. We note that CMB power 
spectra have recently been sensitively measured at multipoles up to $l=10^{4}$ 
with the South Pole Telescope (SPT) and Atacama Cosmology Telescope (ACT) 
high-resolution ground-based telescopes.

The impact of bias induced by uncertainties in modeling the evolution of the LSS 
and its properties is determined from analysis of the matter power spectrum, 
$P(k)$. From Eq.(2.4) the Fisher matrix is
\begin{eqnarray}
F_{ij}=\int_{k_{min}}^{k_{max}}\frac{\partial \ln(P)}
{\partial\lambda_{i}}\frac{\partial \ln(P)}{\partial\lambda_{j}}dN_{k}
\end{eqnarray}
where [10]
\begin{eqnarray}
dN_{k}&=&V_{eff}(k)\frac{2k^{2}dk}{(2\pi)^{2}}\nonumber\\
V_{eff}(k)&=&\int_{z=z_{min}}^{z_{max}}\left(\frac{n(z)P(k,z)}{1+n(z)P(k,z)}
\right)^{2}\frac{dV}{dz}dz,
\end{eqnarray}
and $k_{max}$ and $k_{min}$ mark the smallest and the largest scale in the survey. 
The number of modes sampled by the probe is set by $k_{max}$, and is roughly 
$\sim \frac{4}{3}k_{max}^{3}\frac{V_{eff}}{(2\pi)^{3}}$. The analog of Eq.(2.18) 
in this case is 
\begin{eqnarray}
\sigma_{\lambda}&\sim &\left[N\left(\frac{\partial P/\partial\lambda}{P}
\right)^{2}\right]^{-\frac{1}{2}}\nonumber\\
\frac{\delta_{\lambda}}{\sigma_{\lambda}}&\sim &N^{1/2}\frac{\Delta P}{P}.
\end{eqnarray}

Combining the results from Eqs.(2.18) \& (2.21) we arrive at the basic 
requirement that the fractional error in either the angular or matter power 
spectrum should satisfy
\begin{eqnarray}
\frac{|\Delta C_{l}|}{C_{l}}&\lesssim &\frac{1}{\sqrt{N}}\nonumber\\
\frac{|\Delta P(k)|}{P(k)}&\lesssim &\frac{1}{\sqrt{N}}
\end{eqnarray}
for each $l$ or $k$, namely that the fractional model error in 
the power spectra should be smaller than the square root of the number 
of modes. (Note that the total number of modes in a CMB experiment scales 
as $\sim l_{max}^{2}$, e.g. Eq. 2.13.) 

It is a standard practice in the literature to show that the power spectra 
of systematics, foreground residuals, modeling errors, etc., are suppressed 
to below the cosmic variance level. This is warranted by a marginalization 
procedure over the systematics that results in what is presumed to be an 
unbiased estimate of the cosmological parameters for a relatively low 
cost of (usually) insignificant increase in the statistical uncertainty 
of the inferred cosmological parameters. However, in doing so it is tacitly 
assumed that the systematics model (or the residual systematics model) 
statistically fluctuates around the exact model; this assumption is rarely 
the case: Significant variation between theoretical models for statistical 
measures of the SZ effect is a relevant illustrative example. Predictions 
from these models are never found to fluctuate around each other. Rather, 
for virtually any two models for the SZ power spectrum (normalized to the 
same level at a given scale, e.g. $l=3000$) one typically finds that one 
of the models overestimates the power on small scales while the other 
overestimates it on larger scales. In other words, two different SZ models 
will typically have a different shape in multipole space, effectively 
undermining the basic assumption behind the marginalization procedure. 
If this marginalization path is nevertheless adopted, it would lead to 
an unrealistic level of uncertainty (which is derived in Appendix A),
\begin{eqnarray}
\frac{\delta C_{l}^{sys}}{C_{l}}&\lesssim &N^{-1/4}\nonumber\\
\frac{\delta P(k)^{sys}}{P(k)}&\lesssim &N^{-1/3},
\end{eqnarray}
bounds that are weaker than the bias-free parameter inference requirements, 
Eq.(2.22). The reason for this is that in deriving Eqs.(2.22) we consider 
only that part of systematics, foregrounds, or modeling errors, that 
{\it systematically} increases or decreases the total power spectrum, in 
contrast to Eq.(2.23) which is obtained by assuming that all these systematics 
contribute power which fluctuates around the exact power spectrum.

\section{CMB and LSS Precision Requirements}

The above general assessment of the bias has important implications for the 
realistic degree of precision that can be attained in CMB (2D) and LSS (3D) 
probes, as we now demonstrate. We consider the CMB as a representative for 
probes that are based on the angular power spectrum. Similar probes that will 
not be discussed here are weak gravitational lensing, and redshifted 21-cm 
analyses based on angular power spectra. 

\subsection{CMB Probes}

As previously discussed by Seljak et al. [1], the required numerical precision 
of CMB Boltzmann codes at a given mode $l$ is $1/\sqrt{l}$ if the numerical 
errors are uncorrelated, i.e. fluctuating in $l$. If, on the other hand, there 
is a systematic (i.e. $l$-correlated) error in the power spectrum calculation 
it must satisfy
\begin{eqnarray}
\frac{|\Delta C_{l}|}{C_{l}}\lesssim\frac{1}{l}
\end{eqnarray}
in order not to bias the parameter inference beyond the statistical error, at 
that given $l$. The argument is simply that of mode counting; in a mode-annulus 
of modulus $l$ and width $\Delta l$ there are $2\pi l\Delta l$ modes (assuming 
statistical isotropy). The statistical error in estimating the angular power 
spectrum is therefore $\delta C_{l}/C_{l}\sim 1/\sqrt{l}$ (assuming no 
mode-correlation). However, assuming $\delta C_{l}$ are correlated, i.e. they 
are systematically lower or higher than the real power spectra, the requirement 
becomes $|\Delta C_{l}|/C_{l}\lesssim 1/l$. This will offset the 
$\delta_{\lambda}/\sigma_{\lambda}\propto N^{1/2}$ dependence of the bias, where 
for the CMB  - and other angular power spectra, such as those used in weak 
lensing shear maps or 21-cm forecast -  the number of modes is $N\sim l^{2}$.

The primordial CMB power spectrum dies off very quickly beyond $l\approx 1000$ 
and assuming a multipole cutoff $l_{max}=3000$ is reasonable. Comparing three 
Boltzmann codes, Seljak et al. [1] find that the $0.1\%$ numerical precision 
is marginally achieved. Lesgourgues [11] illustrated that the CLASS and CAMB 
codes agree at the $0.01\%$ level assuming the same evolutionary history. 
Recently, recombination modules for the CMB Boltzmann codes have been updated 
to include small corrections that resulted in only $\sim 0.1\%-0.2\%$ departures 
between CosmoRec [12-13] and HyRec [14-15], again marginally satisfying the bound, 
Eq.(2.22).

Achieving the goal of percent-level precision in determining the cosmological 
parameters may not be realistic given the various sources of systematics. 
It has recently been shown in [16] that the beam window function of PLANCK 
could be calibrated at the $\sim 0.1\%$ level using a parametric beam model, 
but this degrades by a factor of a few if a non-parametric model is assumed. 
A systematic error higher than this benchmark will necessarily propagate 
into the recovered power spectra and will ultimately bias the inferred values 
of cosmological parameters. In Appendix B we provide a more quantitative 
discussion of the precision level that beam calibration has to satisfy in 
order not to violate Eq.(2.22). 

The multifrequency capability of many CMB experiments will enable relatively 
precise removal of most astrophysical foregrounds, due to their non blackbody 
spectrum. A notable exception is the kinematic Sunyaev-Zeldovich (KSZ) effect 
which is essentially a first order Doppler shift of the CMB temperature, and 
therefore does not alter the blackbody spectrum. The estimated level of the 
KSZ from patchy reionization models is $1.5-3.5\mu K^{2}$ [17]. This range 
reflects the theoretical uncertainty in reionization models; it was obtained 
from studying the impact of $\sim 100$ models on the CMB temperature anisotropy. 
The impact of astrophysical processes, degree of patchiness over the relevant 
redshift range, and various feedback processes, is estimated at the 
$\sim 2-3\mu K^{2}$ level at $l=3000$, e.g. [18-21]. Overall, this range of 
variation of the KSZ power due to modeling uncertainty represents $0.1-10\%$ 
perturbation to the primordial CMB on the relevant multipole range $1000<l<3000$. 
Using the non-gaussianity of the SZ effect to remove this contribution is a 
reasonable possibility, but we are not aware of any quantitative study that 
demonstrates that the residual contribution to the CMB power spectrum will 
satisfy Eq.(2.22). 

Taburet et al. [8] have shown that the thermal SZ effect, induced by hot gas 
in galaxy clusters, will significantly bias a few key cosmological parameters, 
even if the most luminous galaxy clusters detected by PLANCK are masked. This 
is due to the rather significant contribution made by the undetected clusters 
to the CMB temperature power spectrum and the finite number of frequency bands, 
instrumental noise, and foregrounds that limit the mass and redshift of 
detectable clusters. While this might not necessarily pose a significant 
challenge to PLANCK science, because cluster masking can benefit from non-CMB 
surveys that are projected to detect $\sim O(10^{5})$ clusters, it has recently 
been shown that warm filamentary structures may introduce comparable bias in 
the inferred cosmological parameters [9]. 

More generally, it remains to be demonstrated that the precision of foreground 
removal techniques can realistically attain the required level implied by 
Eq.(2.22). In the specific case of the thermal SZ effect in clusters, it should 
be emphasized that calculations of the power spectra do not agree at even the 
few percent level due to the highly model-dependent nature of the effect [22-24]. 
This applies not only to the amplitude but, more importantantly, to the 
$l$-dependence of the SZ power spectrum. For example, Millea et al. [25] 
considered the possibility of mitigating the foregrounds-induced bias of 
inferred cosmological parameters by representing the foregrounds by 17 parameters 
which resulted in only $\lesssim 20\%$ increase in the nominal uncertainty. 
While this result is encouraging, in order to assess the reliability of the 
inferred cosmological parameters from PLANCK, compelling evidence has to be 
provided that parametrizing the uncertainty in astrophysical foregrounds with 
only 17 parameters {\it fully} captures the $l$-dependence at the required 
$|\Delta C_{l}|/C_{l}\lesssim 1/l$ level. As we show in Appendix C, allowing 
nuisance free parameters in a `wrong' model does not generally guarantee a 
bias-free cosmological parameter inference.

\subsection{LSS Surveys}

3D surveys that target the matter power spectrum, $P(k)$, clearly probe a 
larger number of modes than the 2D CMB angular power spectrum. The number of 
modes is $\sim \frac{4}{3}k_{max}^{3}\frac{V_{eff}}{(2\pi)^{3}}$, where 
(as specified in the previous section) $V_{eff}$ and $k_{max}$ are the 
effective survey volume and maximum wave number, respectively (Eqs. 2.20). 
For a $\sim 1\, Gpc^{3}$ survey, and $k_{max}\sim 0.1 Mpc^{-1}$ the total 
number of modes is $N\sim 5300$, and the required precision on the matter 
power spectrum is $|\Delta P(k)|/P(k)\lesssim 1.3\%$. The SDSS Lumonius 
Red Galaxy (LRG) survey used $42,000$ modes [26] in the $0.01h/Mpc<k<0.1h/Mpc$ 
range. Reid et al. [27] probed deeper into the quasi-linear regime and used 
data up to $k<0.2h/Mpc$, which resulted in increasing the mode number by a 
factor of $\sim 8$, thereby significantly improving the constraints on the 
cosmological parameters. All these galaxies are observed at $z<1$; how well 
do we know the matter power spectrum at the quasi-linear regime ? 
Calculation of the evolution of $P(k;z)$ can be done perturbatively. 
Two-loop corrections still contribute at the $\sim 1\%$ level at 
quasi-linear scales, e.g. [28]. Going to higher order in regularized 
perturbation theory entails performing integrations at $D=3n-1$ dimensions 
where $n$ is the loop order, e.g. [29]. This is expected to be computationally 
quite prohibitive if the desired numerical accuracy, Eq.(2.22), is to be achieved. 
For example, on scales where 3rd order perturbation terms are non-negligible, 
8D integrations should be done that have to be precise at the $10^{-3}-10^{-7}$ 
level, say, and this will have to be repeated for each parameter set in the 
multi-dimensional parameter-space search for the best-fit cosmological model. 
It is unclear whether this is realistic.

An alternative is to employ numerical simulations; carefully choosing initial 
conditions, time steps, sufficient mass resolution and large volumes will allow 
reaching the rather impressive 1\% accuracy at $k\sim 1 Mpc$ [30], but this 
without including the effect of baryons which in itself is expected to contribute 
at the percent level on these small scales. Also, to be useful for Monte-Carlo 
cosmological parameter search, these calculations of the nonlinear matter power 
spectrum have to be very fast. Running the simulations for only a few cases and 
interpolating between parameter values might result in errors larger than those 
allowed for unbiased cosmological parameters. In addition, a comparison of these 
simulations with CAMB's HALOFIT reveals a 5-10\% discrepancy [30]. 

Another possible avenue is to employ Artificial Neural Networks (ANN) for a fast 
calculation of the matter power spectrum. Agarwal et al. [31] claim to have 
reached the $\lesssim 1\%$ precision on scales $k\leq0.7\,h Mpc^{-1}$ and at 
redshifts $z<2$. However, these neural networks have been trained with HALOFIT, 
which is itself discrepant with [30] at the few percent level.

In parameter estimation forecasts it is customary to adopt the nonlinear-scale 
cutoff at scales where the mass fluctuation inside mass spheres of radius $R$, is 
of order unity: $\sigma(R)\equiv\sqrt{\langle(\frac{\delta M}{M})^{2}\rangle_{R}}
\approx 0.5$, e.g. [32-33]. While this definition of nonlinear scale is intuitive, 
parameter bias is completely unaccounted for. 
In fact, it turns out to overly under-estimate the impact of our ignorance of the 
nonlinear matter power spectrum on the bias of inferred cosmological parameters. In 
their parameter estimation forecast for {\it post-reionization} redshifted-21-cm 
surveys, Visbal, Loeb, \& Wyithe [34], determined the largest mode $k_{max}$ by 
comparing the nonlinear matter power spectrum from HALOFIT to the linear power 
spectrum at the various redshifts they considered. They defined $k_{max}$ at the 
scale where the nonlinear deviates from the linear power spectrum at 10\%. They 
also explored the robustness of their forecast to varying this criterion in the 
range 5-25\% and indeed their analysis shows that the parameter uncertainties 
degrade as $k_{max}^{-3/2}$ (in the range where cosmic variance dominates over 
instrumental noise). Since their 21-cm observation is volume-limited, the number 
of modes can be readily calculated, $N\sim 6.75\times 10^{11}k_{max}^{3}$, where 
$k_{max}$ is in $Mpc^{-1}$ units. Even if we take their most stringent 
$k_{max}=0.1 Mpc^{-1}$, we obtain that for Eq.(2.22) to be satisfied one must know 
the matter power spectrum to better than one part in $10^{4}$. This implies that 
realizing the potential of future post-reionization redshifted-21-cm observations 
seems unlikely. Given the current level of precision in calculations of the matter 
power spectrum, the number of modes will have to be drastically reduced; this will 
result in a significant weakening the stated scientific yield of these probes, 
removing their competitive advantage over other cosmological probes.  

The tantalizing merits of the pre-reionization 21-cm at very high redshifts, and 
its statistical power to constraining cosmology, have been first advocated by 
Loeb \& Zaldarriaga [35]. One may contemplate that {\it pre-reionization} 21-cm at 
high-redshifts is immune to power spectrum non-linearity, which is indeed the case 
above some redshift-dependent sub-Mpc scale. However, use of data down to the baryon 
Jeans scales comes with the penalty of incurring a very 
large bias; the number of modes in these observations is estimated to fall in the 
range $10^{14}-10^{16}$, which will require knowing the matter power spectrum at the 
$\sim 10^{-7}-10^{-8}$ precision. Indeed, it has been shown in [36] 
that nonlinear corrections to the matter power spectrum, even at redshifts 
as high as 30 or 50, may contribute at the sub-percent level at sub-Mpc scales. As 
discussed above, while accounting for higher order corrections to the matter power 
spectrum is theoretically possible (even if at 
the cost of significantly slowing down the search in the multidimensional parameter 
space), it is not clear how many such terms should be included in order to reach the 
fantastic $\sim 10^{-7}-10^{-8}$ precision entailed by requiring unbiased parameter 
estimation from all scales down to the baryon Jeans scale. Cutting off the data above 
some scale $k_{max}$ larger than the Jeans scale $k_{J}$ will degrade the statistical 
uncertainty by $(\frac{k_{max}}{k_{J}})^{3/2}$. 

These considerations are especially relevant when assesssing the scientific yields of 
future surveys. For example, Sigurdson \& Cooray [37] considered the possibility of 
delensing the polarized CMB sky with high-redshifted 21-cm tracer of gravitational 
lenses to the level that will allow constraining the energy scale of inflation 
down to $\mathcal{V}^{1/4}\sim 1.1\times 10^{15}$ GeV, equivalent to tensor-to-scalar 
ratio $\mathcal{T/S}\sim 1.0\times 10^{-6}$ with $l_{\max}\sim 10^{5}$. This 
represents $\sim 3$ orders of magnitude tighter constraint than the ideal CMB 
experiment. However, reliably inferring the energy scale of inflation using this 
method requires a $\sim 10^{-5}$ precision of the theoretical 21-cm anisotropy model, 
a goal that we believe has not been demonstrated to be realistic. In a recent similar 
work, using $l_{\max}\sim 10^{7}$, Book, Kamionkowski \& Schmidt [38] claim that the 
signature of inflationary primordial gravitational waves with as small as 
$\mathcal{T/S}\sim 1.0\times 10^{-9}$ could be detected with future high-redshifted 
21-cm observations. This further boosts the model precision requirement to one part 
in $10^{7}$.

It is well appreciated that realizing the potential of the 21-cm 
probe will be extremely challenging given that foregrounds are expected to be 
$\sim$5 orders of magnitude larger than the 21-cm signal. In Fisher matrix forecasts 
of the science yield of these 21-cm observations, it is common to model the 
frequency-dependence and spatial correlations 
of these foregrounds and marginalize over the model free parameters, e.g. [2], [4], [39-42]. 
These models are often extrapolated from other frequency 
regimes, or are otherwise only partially physically motivated, and are often proposed 
largely due to their functional simplicity.
However, for unbiased cosmological parameter estimation the marginalization process 
only makes sense when the model functional form, i.e. $k$-dependence, 
faithfully captures the shape of the observed power spectrum; the extra freedom 
enabled by adding the nuisance parameters, which are subject to marginalization, 
does not guarantee a bias-free parameter inference. In other words, we generally 
do not expect that using an inadequate model can be {\it fully} compensated for by 
simply marginalizing over free nuisance parameters (as argued in Appendix C).
  
So far we considered only the systematics in the primordial matter power spectrum. In 
reality, the matter power spectrum is determined from observations of dark matter 
biased tracers, e.g., galaxy clustering, galaxy clusters, and perhaps also from redshifted 21-cm 
neutral gas (that follows dark matter halos) in the future. The {\it observed} power 
spectrum is skewed by a scale-dependent multiplicative bias. The observable in galaxy 
surveys is $P_{g}(k;z)=b^{2}(k,z)P(k;z)$, e.g. [43]. The calculation of this 
luminosity-dependent bias is highly non-trivial and model-dependent, so one may 
question its precision. In the case of 21-cm observations at the reionization era, 
we assume the surveys to be only volume-limited; this eases the bias calculation in 
the sense that it is only redshift dependent in that case. 
For far-future high-redshift 21-cm observations at high redshifts, there is very little 
halo bias and in that sense this probe is more immune to this systematic source. 

For the bias $b(z)$ calculation one needs to specify the halo mass function. Until only 
recently the mass function of choice was that of Seth \& Tormen [44-45]. 
This was superseded by the Tinker mass function [46-47], but there are several other 
mass functions. Typically, they are discrepant at more than a few percent on the relevant 
mass--redshift range. In most forthcoming analyses we will only reliably know the redshifts 
of galaxies or galaxy clusters; their mass inference is expected to be very challenging and 
quite biased. When comparing observations to theory a mass function has to be selected in 
order to average the theoretical bias $b(M,z)$ over mass. Also, it is not clear how this 
average should be performed; is it a simple mass average, e.g. [48] ? Is it mass-weighted 
average over redshift, e.g. [34] ? It is therefore clear that uncertainties in mass and 
in the mass function imply the need for arbitrary choices that may easily result in a 
few percent systematic bias in all $k$-modes, ultimately biasing the inferred cosmological 
parameters.

\section{Discussion}

The concordance cosmological model, corroborated by many different cosmological surveys 
that probe the linear and quasi-linear large scale structure, is surprizingly simple on 
the largest scales: Of all concievable universes ours seems to have had 
initial conditions that lead to 
relatively simple LSS, and with properties that are characterized by only a dozen 
cosmological parameters. It now appears that our most important challenge is precision 
measurements of these parameters. The ability to do so depends largely on the sensitivity 
and extent of future cosmological surveys. More specifically, the level of precision 
depends on the susceptibility of the cosmological model to small variations in the key 
parameters, the number of independent (Fourier) modes of survey data, and on the 
degeneracy between the parameters. 

In reality various factors can limit the quality of our deductions from even the most 
precise measurements; almost unavoidably these include some degree of bias in the 
inferred cosmological parameters. In this work we highlighted the quantitative 
`tension' between the statistical error that drops as $N^{-1/2}$ and the dimensionless 
bias that typically grows as $N^{1/2}$, where the number of modes, $N$, increases 
as cosmological experiments become significantly more sensitive. Realizing the 
potential of redshifted 21-cm experiments that will explore $10^{8}-10^{14}$ 
modes requires controlling various systematics, including modeling 
and simulation uncertainties and foreground removal, to the level of 
$O(10^{-4})$-$O(10^{-7})$. Put differently, whereas a model bias of
$0.001\sigma$ could be identified with the CMB satellite WMAP ($\sim
10^{6}$ modes) at $1\sigma$, a bias as small as $10^{-7}\sigma$ would be
discerned with future 21 cm observations ($\sim 10^{14}$ modes); a model
bias larger than this would therefore be at odds with observation and in
any case would bias the inferred cosmological parameters. As emphasized
above, our main concern in this work has been those uncertainties that
systematically shift the power spectra in an unknown fashion (for which we
adopt Eq. (2.22) as the benchmark precision criterion). We stress that
systematics that fluctuate around the exact model can be marginalized over
and typically result in mild degradation of the nominal cosmological
parameter uncertainties without inducing any bias [in which case the
required precision is summarized in Eq. (2.23)]. However, for this to be the
case the model has to exactly capture the shape of the power spectra over
the {\it entire} range of variation of the cosmological parameters. Only
then could the model nuisance parameters be marginalized over and
ensure a bias-free cosmological parameter inference.

With the steadily increasing number of accessible CMB and LSS modes the statistical 
uncertainty in inferred cosmological parameters is ought to improve. However, even a 
slight bias in the theoretical modeling, data analysis, and foreground removal, can 
potentially accrue with increasing number of modes to the level that may possibly bias 
the best-fit cosmological model 
far beyond the nominal statistical uncertainty. This poses a challenge to next 
generation LSS probes, especially to redshifted 21 cm observations which are 
theoretically limited only by the baryon Jeans scale, and are poised to ultimately 
yield measurement database covering a huge number of $\sim 10^{16}$ modes with 
unprecedented capability for precise cosmological parameter inference.

The simple rule of thumb that we highlighted in this work, that the statistical error 
and dimensionless bias on the inferred cosmological parameters scale as $\sim N^{-1/2}$ 
and $\sim N^{1/2}$, respectively, assumes that the cosmological information is entirely 
contained in the angular or matter power spectrum. However, constraining primordial 
non-gaussianity requires working with the angular (matter) bispectrum, in the 2D (3D) 
cases. When the main goal is just to set observational bound on the degree of 
non-gaussianity, without specifiying the non-gaussianity class, then all triangular 
configurations in mode-space are allowed, thereby increasing the number of modes $N$ 
(for a given observational resolution) used in determining the primordial non-gaussianity. 
This only makes the theoretical precision requirements from the model (which is 
contrasted with the data) more demanding.

Our objective in this work has been to highlight a major limitation of upcoming 
cosmological surveys, especially those that are expected to greatly benefit from 
the huge number of modes hitherto unprobed by the current lower resolution and higher 
noise probes. Strictly speaking, our simple arguments apply to either a cosmological 
model with a single free parameter, or to multi-parameter model when the parameters 
are uncorrelated. In practice, this is never the case and cosmological parameters 
do correlate. As the number of usable modes increases there is more information in 
the data to allow breaking the cosmological parameter degeneracies. However, as we 
argue in this work, using a larger number of modes could lead to a stronger bias 
if the model is not sufficiently accurate. To optimize the number of modes used in 
a given cosmological survey the maximal number of modes should be chosen such that 
the statistical error and bias added in quadrature will result in the minimum 
possible error on a given set of parameters.

\section*{Aknowledgements}

We thank Ely Kovetz and Rennan Barkana for their useful comments on an earlier 
version of the draft. We also thank an anonymous referee for useful comments. 
This work was supported by the TAU-UCSD Cosmology Program, 
and by grants from the US-Israel BSF (2008452) and the Israel Science Foundation 
(1496/12).

\appendix

\section{Precision Requirements: Random Modeling Errors}

We outline the rationale behind the standard requirement on precision of theoretical 
power spectra, pointing out explicitly the assumption behind the derivation of 
Eq.(2.23). We do this for the case of angular power spectrum; extending this derivation 
to the case of the matter power spectrum is straightforward. 
Analogous to parameter marginalization one can account for residual foreground 
(or other type of modeling uncertainty) by simply modifying Eq.(2.15)
\begin{eqnarray}
\mathcal{L}'=\int_{\Delta C_{l}}\exp\left[-\sum_{l}\frac{[\Delta C_{l}
+(\frac{\partial C_{l}}{\partial\lambda})(\lambda-\lambda_{0})^{2})
]^{2}}{2(\delta C_{l})^{2}}\right]\exp\left(-\frac{\Delta C_{l}^{2}}{2
\sigma_{C_{l}}^{2}}\right)\mathcal{D}(\Delta C_{l}) ,
\end{eqnarray} 
where we marginalize over the $l$-dependent modeling error of the power spectrum 
(assuming that foregrounds and other systematics have already been accounted for) 
and carry out a functional integration over it with a gaussian prior on the power 
spectrum modeling uncertainty with $\sigma_{C_{l}}$ being the $1-\sigma$ prior 
on the residual modeling error $\Delta C_{l}$. For simplicity we assume it is 
symmetrically distributed around 0, since otherwise this itself would introduce 
a bias. Carrying out the integration one obtains that $\delta C_{l}^{2}$ in 
Eq.(2.15) is replaced by
\begin{eqnarray} 
\delta C_{l}^{2}\rightarrow \delta C_{l}^{2}+\sigma_{C_{l}}^{2}.
\end{eqnarray}
Thus, the inferred parameter is unbiased at the cost of increasing variance. 
As $\delta C_{l}^{2}$ decreases as $l^{-1}$, we expect the residual SZ power 
with $\sigma_{C_{l}}^{2}$ to dominate the denominator (in the above sum) 
at some sufficiently large $l$, resulting in degraded statistical uncertainty of 
several cosmological parameters. This would typically be perceived as a reasonable 
cost when compared to the bias that would otherwise be introduced. 
The first relation in Eq. (2.23) can be obtained from the following consideration: 
If $\delta C_{l}^{2}\propto C_{l}^{2}/l$ then the requirement is $\sigma_{C_{l}}/C_{l}
\lesssim l^{-1/2}$, and recalling that the number of modes $N\sim l_{max}^{2}$ it 
follows that $\delta C_{l}/C_{l}\lesssim N^{-1/4}$. It is straightforward to obtain 
also the second relation in Eqs.(2.23) by employing similar reasoning for the 3D 
matter power spectrum, $P(k)$, and recalling that here $N\sim k_{max}^{3}V$. For 
very large mode numbers, $N$, Eqs.(2.22) \& (2.23) give markedly different 
requirements. In this Appendix we show that making the requirement Eq.(2.23) is 
only warranted under the very strong assumption that the residual systematic power 
spectrum fluctuates around 0. In general, our modeling uncertainty of systematics, 
foreground residual, etc., is hardly of this type; in fact, we rarely know the 
systematic power spectrum up to a fluctuating part. Relevant examples include 
the statistical thermal SZ effect, the KSZ effect from patchy reionization, 
foreground clustering, and beam systematics.

\section{Beam Calibration Requirements}

Features in sky maps on scales smaller than the angular resolution of the telescope 
are effectively smeared by beam dilution. Conventionally, the telescope angular 
response function is modeled as a circular or elliptical gaussian with superimposed 
low order polynomials. By calibrating the beam against a point source such as Mars, 
Jupiter, or Saturn, the best fit model parameters are obtained.

For simplicity, we model the beam dilution effect as a circular gaussian, but the 
result will hold for more general beam models. The {\it measured} power spectrum 
obtained from the map is
\begin{eqnarray}
C_{l}^{measured}=C_{l}^{real}e^{-l^{2}\sigma_{b}^{2}}
\end{eqnarray}
where $\sigma_{b}$ is the gaussian width. Once the beam is calibrated and $\sigma_{b}$ 
is obtained the measured power spectrum is multiplied by $e^{l^{2}\sigma_{b}^{2}}$ 
and the best-fit cosmological model is then obtained from comparing theory and 
observation. In practice, the model chosen for the beam description might not always 
capture all beam features, such as sidelobes, higher order moments beyond the quadrupole 
(parameterized by the beam ellipticity), and non-gaussian features, etc. Choosing 
the `wrong' model might then skew the best-fit procedure. It is conceivable 
that the beam calibration will then result in an effective beam width $\sigma_{b'}$ 
either larger or smaller than $\sigma_{b}$. In this case, the recovered power spectrum 
will be
\begin{eqnarray}
C_{l}^{recovered}=C_{l}^{measured}e^{l^{2}\sigma_{b'}^{2}}
\end{eqnarray}
which is systematically larger or smaller than the actual power spectrum 
$C_{l}^{measured}e^{l^{2}\sigma_{b}^{2}}$. The fractional error in the 
angular power spectrum is therefore
\begin{eqnarray}
\frac{\Delta C_{l}}{C_{l}}=\exp[l^{2}(\sigma_{b}^{2}-\sigma_{b'}^{2})]-1.
\end{eqnarray}
Assuming the difference $\sigma_{b}-\sigma_{b'}$ is small, and combining this 
with Eq.(2.22) results in the condition
\begin{eqnarray}
2l^{2}\sigma_{b}^{2}\mu\lesssim 1/l
\end{eqnarray}
where we defined the beamwidth missmatch $\mu=|\sigma_{b}-\sigma_{b'}|/\sigma_{b}$.
Now, recalling that the maximum multipole probed by the telescope is roughly where 
the beam window function drops to a value $e^{-1}$ times its peak value, we can set 
$l_{max}^{2}\sigma_{b}^{2}\approx 1$ in Eq.(B.4) and obtain
\begin{eqnarray}
2\mu\lesssim 1/l_{max}
\end{eqnarray}
implying that for a beamsize of $\sim 5$ arcminute (the smallest of the PLANCK/HFI 
instruments) the beamwidth has to be calibrated at the $0.1$ arcsecond precision, 
which will be challenging given that point sources (such as planets) morphology in 
the microwave is not known to this very high precision level.

\section{Marginalization and Residual Bias}

Marginalizing over model uncertainties is a standard practice in analyses/forecasts 
of cosmological datasets. This technique is especially relevant for addressing 
{\it residual} foregrounds or systematics. Accounting for these model uncertainties 
is conventionally done by parameterizing systematics and foreground models. In 
most cases these somewhat {\it arbitrary} model choices are motivated by either 
mathematical simplicity or well-understood physics {\it extrapolated} to the regime 
of interest. As mentioned in Section 3 (in our discussion of the CMB and 21-cm surveys), 
this procedure may not meet the precision standards required by unbiased cosmological 
parameter inference, as put forward in this work.

Here we argue that the bias derived in section 2 cannot be simply integrated away by 
{\it assuming} a model for the foregrounds and systematics with free nuisance parameters. 
Therefore, although Fisher matrix analyses of the future performance of marginalization 
techniques applied to CMB and 21-cm observations typically result in only a mild increase 
in the statistical error, this procedure by no means alleviates the bias problem. 

Generalizing Eq.(2.6), we denote the model $\Delta d_{mod}$ for the systematic 
$\Delta d$, so that the likelihood function can now be written as 
\begin{eqnarray}
\mathcal{L}\rightarrow\mathcal{L}'=\exp\left[-\frac{[\Delta d-\Delta d_{mod}
+\frac{d(d)}{d\boldsymbol\lambda}
\cdot(\boldsymbol\lambda-\boldsymbol\lambda_{0})]^{2}}{2(\delta d)^{2}}\right].
\end{eqnarray}
We further parameterize $\Delta d_{mod}=A\Delta\tilde{d}$ with $A$ denoting 
a nuisance model parameter that we marginalize over, assuming a gaussian prior, 
with the dependence on the Fourier mode ($l$ or $k$) included in $\Delta\tilde{d}$ 
\begin{eqnarray}
\mathcal{L}'\rightarrow\mathcal{L}''=\int dA\exp\left[-\frac{[\Delta d-A\Delta\tilde{d}
+\frac{d(d)}{d\boldsymbol\lambda}
\cdot(\boldsymbol\lambda-\boldsymbol\lambda_{0})]^{2}}{2(\delta d)^{2}}-\frac{(A
-A_{0})^{2}}{2\sigma_{A}^{2}}\right]
\end{eqnarray}
where $A_{0}$ and $\sigma_{A}$ characterize the prior range for the nuisance 
parameter $A$. This parameter can, for example, be $A_{SZ}$ of the recent SPT 
and ACT parameterization of the amplitude of the SZ angular power spectrum. 
In this case it is unlikely that N-body simulations or analytic modeling will 
result in an SZ angular power spectrum shape $\Delta\tilde{d}$ identical to 
the actual $\Delta d$ to the required precision $|\Delta C_{l}^{SZ}|/C_{l}
\lesssim 1/l$ up to the maximum $l\sim 3000$ or $4000$ used in the SZ analysis. 
Carrying out the integration over $A$ in Eq.(C.2) in the range $[-\infty,\infty]$ 
is partially justified by assuming that it is known to be several 
$\sigma$ above zero, and partially warranted by the fact that if we set the lower 
integration limit to zero (or a sufficiently small value) then the integration over 
$A$ in an asymmetric range will only weaken the 'power' of marginalization, 
resulting in even a larger bias that what we estimate here. We obtain
\begin{eqnarray}
\mathcal{L}''=\exp\left[-\frac{\beta^{2}}{2}\frac{1}{(\delta d)^{2}
+\sigma_{\alpha}^{2}(\Delta\tilde{d})^{2}}\right]
\end{eqnarray}
where
\begin{eqnarray}
\beta=\Delta d-A_{0}\Delta\tilde{d}+\frac{d(d)}{d\boldsymbol\lambda}
\cdot(\boldsymbol\lambda-\boldsymbol\lambda_{0}).
\end{eqnarray}
Comparing this to Eqs.(2.6) we see that the bias and statistical uncertainty changed
\begin{eqnarray}
\Delta d &\rightarrow &\Delta d-A_{0}\Delta\tilde{d}\nonumber\\
(\delta d)^{2} &\rightarrow & (\delta d)^{2}+\sigma_{\alpha}^{2}
(\Delta\tilde{d})^{2}.
\end{eqnarray}
Typically, $\delta d$ increases by a factor of order unity, e.g. [4], [25], 
and while one wants to use a model where 
$|\Delta d-A_{0}\Delta\tilde{d}|\lesssim 1/\sqrt{N}$, 
this is seldom the case (if at all). In the case of CMB and its SZ foreground, 
different SZ models do not even agree (in amplitude and shape) to better than a few 
tens of percent, far above the $|\Delta d-A_{0}\Delta\tilde{d}|\lesssim 1/l_{max}$ 
requirement.


\begin{thebibliography}{99}

\bibitem{1} Seljak, U., Sugiyama, N., White, M., \& Zaldarriaga, M.
{\it A comparison of cosmological Boltzmann codes: are we ready for high precision 
cosmology?},
{\em PRD} {\bf 68} (2003) 083507
[\href{http://arxiv.org/abs/astro-ph/0306052}{{\tt astro-ph/0306052}}].

\bibitem{2} Santos, M.~G., \& Cooray, A.
{\it Cosmological and Astrophysical Parameter Measurements with 21-cm Anisotropies 
During the Era of Reionization},
{\em PRD} {\bf 74} (2006) 083517
[\href{http://arxiv.org/abs/astro-ph/0605677}{{\tt astro-ph/0605677}}].

\bibitem{3} Pritchard, J.~R., \& Pierpaoli, E.
{\it Constraining massive neutrinos using cosmological 21cm observations},
{\em PRD} {\bf 78} (2008) 065009
[\href{http://lanl.arxiv.org/abs/0805.1920}{{\tt arXiv:0805.1920}}].

\bibitem{4} Mao, Y., Tegmark, M., 
McQuinn, M., Zaldarriaga, M., \& Zahn, O.
{\it How accurately can 21cm tomography constrain cosmology?},
{\em PRD} {\bf 78} (2008) 023529
[\href{http://lanl.arxiv.org/abs/0802.1710}{{\tt arXiv:0802.1710}}].

\bibitem{5} Furlanetto, S.~R., Lidz, A., Loeb, A., et al.
{\it Cosmology from the Highly-Redshifted 21 cm Line},
The Astronomy and Astrophysics Decadal Survey, 2010, 82
[\href{http://lanl.arxiv.org/abs/0902.3259}{{\tt arXiv:0902.3259}}].

\bibitem{6} Santos, M.~G., Cooray, 
A., Haiman, Z., Knox, L., \& Ma, C.-P.
{\it Small-Scale Cosmic Microwave Background Temperature and Polarization 
Anisotropies Due to Patchy Reionization},
{\em ApJ} {\bf 598} (2003) 756-766
[\href{http://arxiv.org/abs/astro-ph/0305471}{{\tt astro-ph/0305471}}].

\bibitem{7} Zahn, O., Zaldarriaga, M., 
Hernquist, L., \& McQuinn, M.
{\it The Influence of Nonuniform Reionization on the CMB},
{\em ApJ} {\bf 630} (2005) 657-666
[\href{http://arxiv.org/abs/astro-ph/0503166}{{\tt astro-ph/0503166}}].

\bibitem{8} Taburet, N., Aghanim, N., Douspis, M., \& Langer, M.\ 2009, 
MNRAS, 392, 1153 (arXiv:0809.1364)
{\it Biases on the cosmological parameters and thermal SZ residuals},
{\em MNRAS} {\bf 392} (2009) 1153-1158
[\href{http://lanl.arxiv.org/abs/0809.1364}{{\tt arXiv:0809.1364}}].

\bibitem{9} Shimon, M., Sadeh, S., \& Rephaeli, Y.
{\it CMB Anisotropy Due to Filamentary Gas: Power Spectrum and Cosmological 
Parameter Bias},
{\em JCAP} {\bf 10} (2012) 38
[\href{http://lanl.arxiv.org/abs/1209.5065}{{\tt arXiv:1209.5065}}].

\bibitem{10} Feldman, H.~A., Kaiser, N., \& Peacock, J.~A.
{\it Power-spectrum analysis of three-dimensional redshift surveys},
{\em ApJ} {\bf 426} (1994) 23-37
[\href{http://arxiv.org/abs/astro-ph/9304022}{{\tt astro-ph/9304022}}].

\bibitem{11} Lesgourgues, J.
{\it The Cosmic Linear Anisotropy Solving System (CLASS) III: Comparision 
with CAMB for LambdaCDM}, (2011)
[\href{http://lanl.arxiv.org/abs/1104.2934}{{\tt arXiv:1104.2934}}].

\bibitem{12} Chluba, J., \& Thomas, R.~M.
{\it Towards a complete treatment of the cosmological recombination problem},
{\em MNRAS} {\bf 412} (2011) 748-764
[\href{http://lanl.arxiv.org/abs/1010.3631}{{\tt arXiv:1010.3631}}].

\bibitem{13} Chluba, J., Fung, J., \& Switzer, E.~R.
{\it Radiative transfer effects during primordial helium recombination},
{\em MNRAS} {\bf 423} (2012) 3227-3242
[\href{http://lanl.arxiv.org/abs/1110.0247}{{\tt arXiv:1110.0247}}].

\bibitem{14} Ali-Ha{\"i}moud, Y., \& Hirata, C.~M.
{\it Radiative transfer effects in primordial hydrogen recombination}, 
{\em PRD} {\bf 82} (2010) 063521 
[\href{http://lanl.arxiv.org/abs/1009.4697}{{\tt arXiv:1009.4697}}].

\bibitem{15} Ali-Ha{\"i}moud, Y., \& Hirata, C.~M.
{\it HyRec: A fast and highly accurate primordial hydrogen and helium 
recombination code}, 
{\em PRD} {\bf 83} (2011) 043513
[\href{http://lanl.arxiv.org/abs/1011.3758}{{\tt arXiv:1011.3758}}].

\bibitem{16} Huffenberger, K.~M., Crill, B.~P., Lange, A.~E., G{\'o}rski, K.~M., 
\& Lawrence, C.~R.
{\it Measuring Planck beams with planets},
{\em AAP} {\bf 510} (2010) A58
[\href{http://lanl.arxiv.org/abs/1007.3468}{{\tt arXiv:1007.3468}}].

\bibitem{17} Mesinger, A., McQuinn, M., \& Spergel, D.~N.
{\it The kinetic Sunyaev-Zel'dovich signal from inhomogeneous reionization: 
a parameter space study},
{\em MNRAS} {\bf 422} (2012) 1403-1417
[\href{http://lanl.arxiv.org/abs/1112.1820}{{\tt arXiv:1112.1820}}].

\bibitem{18} Zhang, P., Pen, U.-L., \& Trac, H.\ 2004, MNRAS, 347, 1224 
(astro-ph/0304534)
{\it Precision era of the kinetic Sunyaev-Zel'dovich effect: 
simulations, analytical models 
and observations and the power to constrain reionization},
{\em MNRAS} {\bf 347} (2004) 1224-1233
[\href{http://arxiv.org/abs/astro-ph/0304534}{{\tt astro-ph/0304534}}].

\bibitem{19} McQuinn, M., Furlanetto, S.~R., Hernquist, L., Zahn, O., 
\& Zaldarriaga, M.
{\it The Kinetic Sunyaev-Zel'dovich Effect from Reionization},
{\em ApJ} {\bf 630} (2005) 643-656
[\href{http://arxiv.org/abs/astro-ph/0504189}{{\tt astro-ph/0504189}}].

\bibitem{20} Shaw, L.~D., Rudd, D.~H., \& Nagai, D. 
{\it Deconstructing the Kinetic SZ Power Spectrum},
{\em ApJ} {\bf 756} (2012) 15
[\href{http://lanl.arxiv.org/abs/0907.1659}{{\tt arXiv:0907.1659}}].

\bibitem{21} Battaglia, N., 
Natarajan, A., Trac, H., Cen, R., \& Loeb, A.
{\it Reionization on Large Scales III: Predictions for Low-ell Cosmic Microwave 
Background Polarization and High-ell Kinetic Sunyaev-Zel'dovich Observables},
[\href{http://lanl.arxiv.org/abs/1211.2832}{{\tt arXiv:1211.2832}}].

\bibitem{22} Shaw, L.~D., Nagai, D., Bhattacharya, S., \& Lau, E.~T. 
{\it Impact of Cluster Physics on the Sunyaev-Zel'dovich Power Spectrum},
{\em ApJ} {\bf 725} (2010) 1452-1465
[\href{http://lanl.arxiv.org/abs/0907.1659}{{\tt arXiv:0907.1659}}].

\bibitem{23} Trac, H., Bode, P., \& Ostriker, J.~P.
{\it Templates for the Sunyaev-Zel'dovich Angular Power Spectrum},
{\em ApJ} {\bf 727} (2011) 94 
[\href{http://lanl.arxiv.org/abs/1006.2828}{{\tt arXiv:1006.2828}}].

\bibitem{24} Zahn, O., Mesinger, A., McQuinn, M., et al.
{\it Comparison of reionization models: radiative transfer simulations and 
approximate, seminumeric models},
{\em MNRAS} {\bf 414} (2011) 727-738
[\href{http://lanl.arxiv.org/abs/1003.3455}{{\tt arXiv:1003.3455}}]. 

\bibitem{25} Millea, M., Dor{\'e}, O., Dudley, J., et al.
{\it Modeling Extragalactic Foregrounds and Secondaries for Unbiased 
Estimation of Cosmological 
Parameters from Primary Cosmic Microwave Background Anisotropy},
{\em ApJ} {\bf 746} (2012) 4
[\href{http://lanl.arxiv.org/abs/1102.5195}{{\tt arXiv:1102.5195}}].

\bibitem{26} Tegmark, M., Eisenstein, D.~J., Strauss, M.~A., et al.
{\it Cosmological constraints from the SDSS luminous red galaxies},
{\em PRD} {\bf 74} (2006) 123507
[\href{http://arxiv.org/abs/astro-ph/0608632}{{\tt astro-ph/0608632}}].

\bibitem{27} Reid, B.~A., Percival, W.~J., Eisenstein, D.~J., et al.
{\it Cosmological constraints from the clustering of the Sloan Digital 
Sky Survey DR7 luminous red galaxies},
{\em MNRAS} {\bf 404} (2010) 60-85
[\href{http://lanl.arxiv.org/abs/0907.1659}{{\tt arXiv:0907.1659}}].

\bibitem{28} Taruya, A., Bernardeau, 
F., Nishimichi, T., \& Codis, S.\ 2012 (arXiv:1208.1191) 
{\it RegPT: Direct and fast calculation of regularized cosmological power 
spectrum at two-loop order},
(2012)
[\href{http://lanl.arxiv.org/abs/1208.1191}{{\tt arXiv:1208.1191}}].

\bibitem{29} Shoji, M., \& Komatsu, E.
{\it Third-Order Perturbation Theory with Nonlinear Pressure},
{\em ApJ} {\bf 700} (2009) 705-719
[\href{http://lanl.arxiv.org/abs/0903.2669}{{\tt arXiv:0903.2669}}].

\bibitem{30} Heitmann, K., White, M., Wagner, C., Habib, S., \& Higdon, D.
{\it The Coyote Universe. I. Precision Determination of the Nonlinear Matter 
Power Spectrum},
{\em ApJ} {\bf 715} (2010) 104-121
[\href{http://lanl.arxiv.org/abs/0812.1052}{{\tt arXiv:0812.1052}}].

\bibitem{31} Agarwal, S., Abdalla, 
F.~B., Feldman, H.~A., Lahav, O., \& Thomas, S.~A.
{\it PkANN - I. Non-linear matter power spectrum interpolation through 
artificial neural networks},
{\em MNRAS} {\bf 424} (2012) 1409-1418,
[\href{http://lanl.arxiv.org/abs/1203.1695}{{\tt arXiv:1203.1695}}].

\bibitem{32} Seo, H.-J., \& Eisenstein, D.~J.
{\it Probing Dark Energy with Baryonic Acoustic Oscillations from Future 
Large Galaxy Redshift Surveys},
{\em ApJ} {\bf 598} (2003) 720-740
[\href{http://arxiv.org/abs/astro-ph/0307460}{{\tt astro-ph/0307460}}].

\bibitem{33} Carbone, C., Verde, L., Wang, Y., \& Cimatti, A.
{\it Neutrino constraints from future nearly all-sky spectroscopic galaxy surveys},
{\em JCAP} {\bf 3} (2011) 30
[\href{http://lanl.arxiv.org/abs/1012.2868}{{\tt arXiv:1012.2868}}].

\bibitem{34} Visbal, E., Loeb, A., \& Wyithe, S.\ 2009, JCAP, 10, 30 (arXiv:0812.0419)
{\it Cosmological constraints from 21cm surveys after reionization},
{\em JCAP} {\bf 10} (2009) 30
[\href{http://lanl.arxiv.org/abs/0812.0419}{{\tt arXiv:0812.0419}}]. 

\bibitem{35} Loeb, A., \& Zaldarriaga, M.
{\it Measuring the Small-Scale Power Spectrum of Cosmic Density Fluctuations through 
21cm Tomography Prior to the Epoch of Structure Formation},
{\em PRL} {\bf 92} (2004) 211301
[\href{http://arxiv.org/abs/astro-ph/0312134}{{\tt astro-ph/0312134}}].

\bibitem{36} Lewis, A., \& Challinor, A.
{\it 21cm angular-power spectrum from the dark ages},
{\em PRD} {\bf 76} (2007) 083005
[\href{http://arxiv.org/abs/astro-ph/0702600}{{\tt astro-ph/0702600}}].

\bibitem{37} Sigurdson, K., \& Cooray, A.
{\it Cosmic 21 cm Delensing of Microwave Background Polarization and the Minimum 
Detectable Energy Scale of Inflation},
{\em PRL} {\bf 95} (2005) 211303
[\href{http://arxiv.org/abs/astro-ph/0502549}{{\tt astro-ph/0502549}}].

\bibitem{38} Book, L., Kamionkowski, M., \& Schmidt, F.
{\it Lensing of 21-cm Fluctuations by Primordial Gravitational Waves},
{\em PRL} {\bf 108} (2012) 211301
[\href{http://lanl.arxiv.org/abs/1112.0567}{{\tt arXiv:1112.0567}}].

\bibitem{39} Santos, M.~G., Cooray, A., \& Knox, L.
{\it Multifrequency analysis of 21 cm fluctuations from the Era of Reionization},
{\em ApJ} {\bf 625} (2005) 575-587
[\href{http://arxiv.org/abs/astro-ph/0408515}{{\tt astro-ph/0408515}}].

\bibitem{40} McQuinn, M., Zahn, O., 
Zaldarriaga, M., Hernquist, L., \& Furlanetto, S.~R.
{\it Cosmological Parameter Estimation Using 21 cm Radiation from the Epoch 
of Reionization},
{\em ApJ} {\bf 653} (2006) 815-834
[\href{http://arxiv.org/abs/astro-ph/0512263}{{\tt astro-ph/0512263}}].

\bibitem{41} Wang, X., Tegmark, M., Santos, M.~G., \& Knox, L.
{\it 21 cm Tomography with Foregrounds},
{\em ApJ} {\bf 650} (2006) 529-537
[\href{http://arxiv.org/abs/astro-ph/0501081}{{\tt astro-ph/0501081}}].

\bibitem{42} Bowman, J.~D., Morales, M.~F., \& Hewitt, J.~N.
{\it Constraints on Fundamental Cosmological Parameters with Upcoming 
Redshifted 21 cm Observations},
{\em ApJ} {\bf 661} (2007) 1-9
[\href{http://arxiv.org/abs/astro-ph/0512262}{{\tt astro-ph/0512262}}].

\bibitem{43} Tegmark, M., Blanton, M.~R., Strauss, M.~A., et al.
{\it The Three-Dimensional Power Spectrum of Galaxies from the Sloan 
Digital Sky Survey},
{\em ApJ} {\bf 606} (2004) 702-740
[\href{http://arxiv.org/abs/astro-ph/0310725}{{\tt astro-ph/0310725}}].

\bibitem{44} Sheth, R.~K., \& Tormen, G.
{\it Large scale bias and the peak background split},
{\em MNRAS} {\bf 308} (1999) 119-126
[\href{http://arxiv.org/abs/astro-ph/9901122}{{\tt astro-ph/9901122}}].

\bibitem{45} Sheth, R.~K., Mo, H.~J., \& Tormen, G.
{\it Ellipsoidal collapse and an improved model for the number and 
spatial distribution of dark matter haloes},
{\em MNRAS} {\bf 323} (2001) 1-12
[\href{http://arxiv.org/abs/astro-ph/9907024}{{\tt astro-ph/9907024}}].

\bibitem{46} Tinker, J., Kravtsov, A.~V., Klypin, A., et al.
{\it Toward a Halo Mass Function for Precision Cosmology: 
The Limits of Universality},
{\em ApJ} {\bf 688} (2008) 709-728
[\href{http://lanl.arxiv.org/abs/0803.2706}{{\tt arXiv:0803.2706}}].

\bibitem{47} Tinker, J.~L., 
Robertson, B.~E., Kravtsov, A.~V., et al.
{\it The Large-scale Bias of Dark Matter Halos: 
Numerical Calibration and Model Tests},
{\em ApJ} {\bf 724} (2010) 878-886
[\href{http://lanl.arxiv.org/abs/1001.3162}{{\tt arXiv:1001.3162}}].

\bibitem{48} Wang, S., Khoury, J., 
Haiman, Z., \& May, M.\ 2004, PRD, 70, 123008 (astro-ph/0406331)
{\it Constraining the evolution of dark energy with a combination 
of galaxy cluster observables},
{\em PRD} {\bf 70} (2004) 123008
[\href{http://arxiv.org/abs/astro-ph/0406331}{{\tt astro-ph/0406331}}].

\end{thebibliography}
\end{document}